\documentstyle[12pt]{article}       
 \begin{document}
 

\title{On the solution of the EPR paradox and the explanation of
the violation of Bell's inequality} 
\author{Gyula Bene\\
Institute for Solid State Physics, E\"otv\"os University\\
M\'uzeum k\"or\'ut 6-8, H-1088 Budapest, Hungary\\
tel/fax: +36 1 266 7509, E-mail: bene@sph.elte.hu}

\date{\today}
\maketitle

\begin{abstract}
A new theory is proposed offering a 
consistent conceptual basis for nonrelativistic
quantum  mechanics. The Einstein-Podolsky-Rosen (EPR) paradox is solved
and the violation of Bell's inequality is explained
 by maintaining realism, inductive inference
and Einstein separability.  
\end{abstract}

\section{ Introduction}

Quantum theory is obviously the greatest achievement 
in theoretical physics of the 20th century. 
It has been successfully applied to the description of
electrons in atoms, molecules and solids, of nuclei,
radiation, elementary particles - an enormous number
of different phenomena. Nevertheless, it is astonishing
that during all these successes the basic concepts,
especially the measurement and the collapse of the wave
function \cite{von Neumann} (when we confine our considerations to
nonrelativistic quantum mechanics only) has remained
unclear and controversial, essentially no progress has been
achieved for almost 70 years, only the confusion
has increased. This unprecedental situation has lead
to an irrational 'folklore' in the physics community:
most physicists settle for themselves the deep and tantalizing
fundamental problems of quantum mechanics by telling
that all this is philosophy (and this is not 
meant to be a compliment), and for
all practical purposes standard quantum mechanics is
completely satisfactory. Indeed, the last statement is true
as long as one compares experimental cross-sections,
transition probabilities or spectral wave numbers with
the theoretical predictions. Somehow this part of the theory is
not sensitive to the logical consistency of the
basic concepts.\footnote{Note that this is certainly not the case
in quantum cosmology. One important physical motivation
for studying the fundamental questions of quantum mechanics is
the hope that the correct answer will help establishing
the quantum theory of gravitation.} Nevertheless, 
the fundamental problems of quantum mechanics cannot
be considered as mere philosophical questions, as two famous
results, the EPR paradox\cite{EPR} and 
the violation of Bell's inequality\cite{hid}
clearly demonstrate. Both of these seem to imply that
quantum mechanics violates Einstein separability\footnote{At
least if one assumes that quantum mechanics is a {\em complete} 
theory (in case of the EPR paradox) and if one maintains {\em
realism} and {\em inductive inference}\cite{Esp} (in case of the 
violation of Bell's inequality).}, an obvious
physical requirement which follows from the principle of locality
and means that separated systems (i.e., which are prevented
from any interaction with each other) cannot
influence each other. When a physical theory violates a well established, 
basic physical principle, that is certainly not a philosophical
problem.  

In order to look at the problem more closely 
let us briefly recall the essence of the two results 
mentioned above. The EPR paradox draws the fact to our attention 
that if we are given two separated systems in an
entangled state (which is the result of a previous interaction), and 
we perform a measurement on one of the systems, then, according
to the usual rules of quantum mechanics, the state of the {\em other,
undisturbed} system will depend on what a quantity has been
measured. This dependence is so strong, that one may end up with
such states for the undisturbed system (in case of different
measurements), that are eigenstates of {\em noncommuting} operators. 

The violation of Bell's inequality is perhaps even more powerful
as it is experimentally proven\cite{exp}. Bell's inequality
refers to a situation when  measurements are performed on each
of the separated systems mentioned above. One assumes that any
correlation between the results of the measurements performed
on the different systems can come only from the previous
interaction which created the entangled state. Therefore, one supposes that
 there are some
stable properties attached to each system, so that these
properties 'store' the correlation after the systems have become
separated, and they determine (at least in a probabilistic sense) 
the outcome of the corresponding measurements.
With these assumptions one finds that the correlations cannot be
arbitrary but must satisfy a certain inequality. This is Bell's
inequality. The correlations may be calculated quantum
mechanically, and the quantum prediction {\em does not} always 
satisfy Bell's inequality. Correlations are measurable
quantities, and experiments have proved the correctness of 
the  quantum prediction and thus the violation of Bell's
inequality. 

What does it mean? Most people seem to believe that the above
results indeed imply that separated systems can influence each
other. Nevertheless, we maintain that such a conclusion is 
physically unacceptable. The principle of locality (or Einstein
separability) has served us well in every branch of physics,
even in quantum physics, including the most sophisticated
quantum field theories. It is rather hard to believe that it
would fail only in case of measurements. After all, a
measurement is just an interaction between two physical 
systems, one of them being a macroscopic measuring device
which consists of atoms whose structure and interactions
are rather well understood in terms of quantum mechanics.
There is no room for a mysterious nonlocal influence.

In case of the EPR paradox it is obvious that if one wants to
maintain Einstein separability, he must change the
interpretation and must replace the collapse of the wave
function with something else. 

The experimentally observed violation of Bell's inequality
is more puzzling. The derivation of Bell's inequality
is completely independent of quantum mechanics, it is
based on a few very fundamental assumptions\cite{Esp}: realism,
inductive inference and Einstein separability. Realism and
inductive inference are not less important in physics than 
Einstein separability, so we do not want to give up them,
either. The only way out can be if there is some further,
independent and hidden assumption, which seems to us 
obvious, but which is not valid in quantum mechanics.

In the present letter it will be shown that it is indeed the
case.  One may reinterpret the meaning and the interrelations of the
quantum states such a manner that the EPR paradox 
and the violation of Bell's inequality gain a natural
explanation without giving up realism, inductive inference or
Einstein separability. The hidden, not allowed assumption mentioned above
is connected to the fact that in the new theory 
the simultaneous existence of the different states is usually restricted
in a particular way, namely, although each state exists separately, 
they cannot be compared without essentially disturbing the system.
In case of the violation of Bell's inequality it turns out that
the states of the measuring 
devices and those which 'store' the correlations are not
comparable (as any attempt for a comparison changes the correlations),
so the usual picture about stable properties 
which are comparable at any time with anything
and 'store' the correlations does not apply,
although the correlations may be attributed exclusively
to the 'common past' (previous interaction) of the particles.

As we shall see, the theory proposed here involves a rather
fundamental change of the basic concepts of quantum mechanics.
Measurements will not be the primary concepts any longer,
there will be no collapse of the wave function, and quantum
states exhibit a new kind of dependence on {\em quantum
reference systems}, a new concept to be explained below.
On the other hand, the quantum mechanics of closed systems
does not change, especially, Schr\"odinger's equation
remains to be valid, and thus all the usual measurable
consequences are unchanged. The {\em only} advantage of the
present theory is that it respects the principle of locality
and offers a consistent and physically acceptable basis for
quantum mechanics.

\section{The basic new concept: quantum reference systems}

Let us consider a simple example, namely, an idealized measurement
of an $\hat S_z$ spin component of a spin-$\frac{1}{2}$ 
particle. Be the particle $P$ initially in the state
\begin{eqnarray}
\alpha |\uparrow>+\beta|\downarrow>\;,\label{u1}
\end{eqnarray}
where $|\alpha|^2+|\beta|^2=1$ and the states $|\uparrow>$ and $|\downarrow>$ are the
eigenstates of $\hat S_z$ corresponding to the eigenvalues
$\frac{\hbar}{2}$ and $-\frac{\hbar}{2}$, respectively.
The other quantum numbers and variables have been suppressed. The dynamics of the
measurement is given by the relations
$|\uparrow>|m_0>\; \rightarrow \; |\uparrow>|m_{\uparrow}>$ and  
$|\downarrow>|m_0>\; \rightarrow \; |\downarrow>
|m_{\downarrow}>$, where $|m_0>$ stands for the state 
of the measuring device $M$ (e.g. a Stern-Gerlach apparatus) 
before the
measurement (no spot on the photographic plate),
while $|m_{\uparrow}> $ ($|m_{\downarrow}>$) 
is the state of the measuring device after the measurement
that corresponds
to the measured spin value $\frac{\hbar}{2}$ ($-\frac{\hbar}{2}$).
The shorthand notation $\rightarrow$ stands 
for the unitary time evolution during
the measurement, which is assumed to 
fulfill the time dependent Schr\"odinger equation 
corresponding to the total Hamiltonian of the combined 
$P+M$ system. As the initial state of the particle is
given by Eq.(\ref{u1}), the linearity of the Schr\"odinger equation
implies that the measurement process can be 
written as
\begin{eqnarray}
(\alpha |\uparrow>+\beta |\downarrow>)|m_0> 
\;\rightarrow \; |\Psi>=\alpha |\uparrow>|m_{\uparrow}>
+\beta |\downarrow>|m_{\downarrow}>\quad.
 \label{u2}
\end{eqnarray}
Let us consider now the state of the measuring device $M$ after the measurement. As the
combined system $P+M$ is in an entangled state, the 
measuring device has no own wave function and may be described
by the {\em reduced density matrix}\cite{Landau} 
\begin{eqnarray}
\hat \rho_M=Tr_P\left(|\Psi><\Psi|\right)
=|m_{\uparrow}>|\alpha|^2<m_{\uparrow}|
+ |m_{\downarrow}>|\beta|^2<m_{\downarrow}|\quad,\label{u3}
\end{eqnarray}
where $Tr_P$ stands for the trace operation in the Hilbert
space of the particle $P$. Nevertheless, if we look at the measuring device,
we certainly see that either $\frac{\hbar}{2}$ or $-\frac{\hbar}{2}$ spin component
has been measured, that correspond to the states $|m_{\uparrow}> $ and $|m_{\downarrow}>$,
respectively. These are obviously not the same as the state (\ref{u3}).
Indeed, $|m_{\uparrow}> $ and $|m_{\downarrow}>$ are pure states
while $\hat \rho_M$ is a mixture of them. Why do we get different states?
According to orthodox quantum mechanics,
one may argue as follows. The reduced density matrix $\hat \rho_M$ 
has been calculated from the state $|\Psi>$ (cf. Eq.(\ref{u2}))
of the whole system $P+M$. A state is a result of a measurement
(the preparation), so we may describe $M$ by $\hat \rho_M$ if we have
gained our information about $M$ from a measurement done on $P+M$. 
On the other hand, looking at the measuring device directly
is equivalent with a measurement done directly on $M$.
In this case $M$ is described by either $|m_{\uparrow}> $ or $|m_{\downarrow}>$.
We may conclude that performing  measurements on 
different systems (each containing the system we want to decribe)
gives rise to different descriptions 
(in terms of different states).
Let us call the system which has been measured (it is $P+M$
in the first case and $M$ in the second case) the {\em quantum
reference system}. Using this terminology, we may tell that
we make a measurement on the quantum reference system $R$, thus we prepare
its state $|\psi_R>$ and using this information we calculate
the state $\hat \rho_S(R)=Tr_{R\setminus S} |\psi_R><\psi_R|$ 
of a subsystem $S$. We shall call
$\hat \rho_S(R)$  the state of $S$ with respect to $R$. 
Obviously $\hat \rho_R(R)=|\psi_R><\psi_R|$, thus $|\psi_R>$
may be identified with the state of the system $R$ 
with respect to itself.

Let us emphasize that up to now, despite of the new terminology,
there is nothing new in the discussion. We have merely 
considered some rather elementary consequences of basic quantum mechanics. 

Let us return now to the question why the state of
the system $S$ (i.e., $\hat \rho_S(R)$) depends
 on the choice of the quantum reference
system $R$. In the spirit of the Copenhagen interpretation
one would answer that in quantum mechanics measurements
unavoidably disturb the systems, therefore, if we perform
measurements on  different surroundings $R$, this disturbance is
different, and this is reflected in the $R$-dependence of $\hat \rho_S(R)$.
Nevertheless, this argument is not compelling. We may also
assume that the states of the systems have already existed before the
measurements, and that there may exist measurements which do not change
these states.  Then the $R$-dependence
of $\hat \rho_S(R)$ becomes an inherent property of quantum mechanics.
Let us leave at this decisive point the traditional framework of quantum mechanics
and follow the new line just sketched. 
 
The meaning of the quantum reference systems is now analogous
to the classical coordinate systems. Choosing a
classical coordinate system means that we imagine
what we would experience if we were there. Similarly,
choosing a quantum reference system $R$ means that we
imagine what we would experience if we did a measurement on
$R$ that does not disturbe $\hat \rho_R(R)=|\psi_R><\psi_R|$. 
In order to see that such a measurement exists, consider an
operator $\hat A$ (which acts on the Hilbert space of $R$) whose
eigenstates include $|\psi_R>$. The measurement of $\hat A$
will not disturbe $|\psi_R>$. Let us emphasize that the
possibility of nondisturbing measurements is an expression
of realism: the state $\hat \rho_R(R)$ exists independently
whether we measure it or not.

As the dependence of $\hat \rho_S(R)$ on $R$
is a fundamental property now, one has to specify 
the relation of the different states in  terms of suitable
postulates. Below we list these postulates.

\section{Rules of the new framework}

{\bf Postulate 1. \em The system S to be described is a subsystem of the
reference system R.} 

{\bf Postulate 2. \em The state $\hat \rho_S(R)$ is a positive definite, 
Hermitian
operator with unit trace, acting on the Hilbert space of $S$.}

{\bf Definition 1. \em $\hat \rho_S(S)$ is called the internal state 
of $S$.}

{\bf Postulate 3. \em The internal states $\hat \rho_S(S)$ are always 
projectors, i.e., $\hat \rho_S(S)=|\psi_S><\psi_S|$.}

In the following these projectors will be identified with the 
corresponding wave functions $|\psi_S>$ (as they are uniquely
related, apart from a phase factor).
 
{\bf Postulate 4. \em The state of a system $S$ with respect to the
reference system $R$ (denoted by $\hat \rho_S(R)$) is the reduced
density matrix of $S$ calculated from the internal state of $R$,
i.e.
$\hat \rho_S(R)=Tr_{R\setminus S} \left(\hat \rho_R(R)
\right)$,
where $R\setminus S$ stands for the subsystem of $R$ 
complementer to $S$.}

{\bf Definition 2.\em  An isolated system is 
such a system that has not been interacting 
with the outside world. A closed system
is such a system that is not interacting with any other
system at the given instant of time 
(but might have interacted in the past).}

{\bf Postulate 5.\em  If $I$ is an isolated system then its state is 
independent of the reference system $R$:}
$
\hat \rho_I(R)=\hat \rho_I(I)$.

{\bf Postulate 6.\em  If the reference system $R=I$ is an isolated one 
then the state $\hat \rho_S(I)$ commutes with the
internal state $\hat \rho_S(S)$.}

This means that the internal state of $S$ coincides with
one of the eigenstates of $\hat \rho_S(I)$.

{\bf Definition 3.\em  The possible internal states are the eigenstates 
of
$\hat \rho_S(I)$ provided that the reference system $I$ is an
 isolated one.}
 
{\bf Postulate 7. \em If $I$ is an isolated system, then the
probability $P(S,j)$ that the
eigenstate $|\phi_{S,j}>$ 
of $\hat \rho_S(I)$ coincides with $\hat \rho_S(S)$ 
is given by the corresponding eigenvalue $\lambda_j$.} 

{\bf Postulate 8.\em  The result of a measurement is contained 
unambigously in the internal state of the measuring device.}

{\bf Postulate 9. \em If there are $n$ ($n=2,\;3,\;...$) disjointed physical systems, 
denoted by
\hfill\break
$S_1, S_2, ... S_n$, all contained in the isolated reference 
system $I$ and 
having the 
possible internal states
$|\phi_{S_1,j}>, |\phi_{S_2,j}>,...,|\phi_{S_n,j}>$, respectively, 
then the joint
probability that $|\phi_{S_i,j_i}>$ 
coincides with the internal state of $S_i$ ($i=1,..n$)
is given by
\begin{eqnarray}
P(S_1,j_1,S_2,j_2,...,S_n,j_n)\mbox{\hspace{5cm}}\nonumber\\
=Tr_{S_1+S_2+...+S_n} [\hat \pi_{S_1,j_1} 
\hat \pi_{S_2,j_2}
...\hat \pi_{S_n,j_n}\hat \rho_{S_1+S_2+...+S_n}(I)], \label{u5}
\end{eqnarray}
where $\hat \pi_{S_i,j_i}=|\phi_{S_i,j_i}><\phi_{S_i,j_i}|$.}

{\bf Postulate 10.\em  The internal state $|\psi_C>$ of a closed system 
$C$
satisfies the time dependent Schr\"odinger equation}
$i\hbar \partial_t |\psi_C>=\hat H |\psi_C>$.

Here $\hat H$ stands for the Hamiltonian.
\vskip0.5cm

It is an important feature of the theory that the states defined
with respect to different quantum reference systems are not
necessarily comparable. What does it mean? As we have seen one
may determine the state $\hat \rho_R(R)$ of any system $R$
without disturbing it if one performs a suitable measurement on
$R$. We may do it with {\em one} arbitrarily chosen system $R$.
But can we do this with {\em two} (or more) systems $R_1$, $R_2$
($R_3$, ...) at the same time (or in succession)? In other
terms: may we always attribute a physical meaning to the 
{\em simultaneous} existence of $\hat \rho_{R_1}(R_1)$ and
$\hat \rho_{R_2}(R_2)$? The answer is no. If we perform 
a nondisturbing measurement on $R_1$, this will disturbe
$\hat \rho_{R_2}(R_2)$ except when $R_2\subseteq R_1$ or 
$R_1\cap  R_2=\emptyset$. Let us demonstrate this
phenomenon on a simple example. Consider two different spin-$\frac{1}{2}$ particles
$P_1$ and $P_2$ and be $R_1=P_1$, $R_2=P_1+P_2$.
 Suppose that $P_1+P_2$ is initially an isolated system
and \begin{eqnarray}
|\psi_{P_1+P_2}>=a|1,\uparrow>|2,\downarrow>-b|1,
\downarrow>|2,\uparrow>\quad,\label{u6}
\end{eqnarray}
where $|a|^2+|b|^2=1$. 
The notation $|1,\uparrow>$ stands for such a state of the first
particle, where the $z$ component of the spin (denoted by
$\hat S_{1z}$) has the definite value $+\frac{\hbar}{2}$. The other
notations have an analogous meaning. 
Further, suppose  that $|\psi_{P_1}>=|1,\uparrow>$.
Measuring $\hat S_z$ on $P_1$  we may write 
\begin{eqnarray}
|\psi_{P_1+P_2}>|m_0>\;\rightarrow \;a|1,\uparrow>|2,\downarrow>|m_{\uparrow}>-b|1,
\downarrow>|2,\uparrow>|m_{\downarrow}>\quad,
\label{u7}
\end{eqnarray}
where $|m_0>$, $|m_{\uparrow}>$ and $|m_{\downarrow}>$ are the
states of the measuring device $M$ as defined earlier.
Eq.(\ref{u7}) implies 
\begin{eqnarray}
\hat
\rho_{P_1+P_2}(P_1+P_2+M)=|1,\uparrow>|2,\downarrow>|a|^2<2,\downarrow|<1,\uparrow|
\nonumber\\
\mbox{\hspace{2cm}}+|1,\downarrow>|2,\uparrow>|b|^2<2,\uparrow|<1,\downarrow|
\label{u8}
\end{eqnarray}
thus $|\psi_{P_1+P_2}>$ has changed and has become due to the measurement either 
$|1,\uparrow>|2,\downarrow>$
or
$
|1,\downarrow>|2,\uparrow>$.

We may summarize this situation as follows: all the states $\hat
\rho_S(R)$ exist {\em individually} 
(where $R$ and $S\subset R$ may be any existing
system), as we may choose a system $R$ at will and
may perform a suitable measurement on it in order to learn
the states $\hat \rho_S(R)$ (now $R$ is fixed) without
changing them. Nevertheless, once we perform this
measurement we unavoidably disturb the states $\hat \rho_{\tilde
S}(\tilde R)$ when $\tilde R\not\subseteq R$ and $\tilde R
\cap R \neq \emptyset$, thus preventing us from
learning these latter states. Therefore, no physical meaning may be
attributed to the {\em simultaneous} existence of the states 
$\hat \rho_S(R)$ and $\hat \rho_{\tilde S}(\tilde R)$
although they do exist {\em separately}.

One may perhaps think that the above property questions
the reality of the states or the realism of the theory.
Let us consider a simple classical analogue demonstrating
this is not the case. It is known that general
relativity allows coordinate systems even inside of a black
hole. Consider now two different black holes and
introduce a coordinate system inside each of them.
Probably no one doubts the reality of the descriptions
with respect to these coordinate systems. We may indeed
check what we may experience with respect to {\em one} of these
systems. We may freely choose one of the black holes and may
fall into it. Then we see what is inside. However, if we do so
we cannot come back and thus automatically prevent ourselves
from learning the other blackhole interieur. Therefore, 
similarly to the quantum case, no physical meaning can be
attributed to the {\em simultaneous} existence of descriptions 
with respect to the above two coordinate systems. Of course, we
do not want to say that there is any deeper connection between
the underlying physics of the quantum case and that of the above 
classical example.

Let us mention one more unusual feature of the quantum reference
systems, namely, that the states of the same system with respect
to different quantum reference systems are not uniquely related.
Indeed, in our first example (cf. Eq.(\ref{u3})) 
$
\hat \rho_M(P+M)=|m_{\uparrow}>|\alpha|^2<m_{\uparrow}|
+ |m_{\downarrow}>|\beta|^2<m_{\downarrow}|\quad,
$
while $\hat \rho_M(M)$ can be either $|m_\uparrow>$ (with
probability $|\alpha|^2$) or $|m_\downarrow>$ (with
probability $|\beta|^2$). Certainly this feature is
an expression of the indeterministic nature of quantum
mechanics. 

Let us mention finally that, strictly speaking, the whole
formalism of the present theory is connected to the
experience only via Postulate 8. One can indeed see that 
measurement is not the primary concept any longer. The theory 
works such a way that - on the basis of the results of previous
measurements - one assumes an initial state of an isolated
system which includes the present measuring device as well, 
calculates the final state from the Schr\"odinger equation
and finally, using the Postulates, deduces the state $\hat
\rho_M(M)$ of the measuring device $M$ to get a prediction
concerning the outcome of the measurement. Certainly, this
prediction will be usually probabilistic.

\section{Solution of the EPR paradox}

Consider again the two-particle system $P_1+P_2$ (cf. Eq.(\ref{u6}) ).
Suppose one performs a measurement on the first particle. Let us 
consider the situation
when one measures $\hat S_{1z'}$, where the $z'$ axis is
obtained from the $z$ axis by
a rotation at an angle $\delta$ around the $x$ axis. 
The initial state of the whole
system (including the measuring device) is given by
$
|m_0>  (
a|1,\uparrow>  |2,\downarrow>-b|1,
\downarrow>  |2,\uparrow>)\quad,
$
where $|m_0>$ stands for the initial state of the measuring device.
The time evolution during the measurement 
can be established by using the relations
$
|m_0>  |1,\delta,\uparrow>\;\rightarrow \;|m_+>
  |1,\delta,\uparrow>
$
and
$
|m_0>  |1,\delta,\downarrow>\;\rightarrow \;|m_->
  |1,\delta,\downarrow>
$,
where
$
|1,\delta,\uparrow>=\cos(\frac{\delta}{2})|1,\uparrow>
-\sin(\frac{\delta}{2})|1,\downarrow>\quad,
$
and
$
|1,\delta,\downarrow>=\sin(\frac{\delta}{2})|1,\uparrow>
+\cos(\frac{\delta}{2})|1,\downarrow>\quad.
$
Thus the final state of the whole system is
\begin{eqnarray}
|m_+> |1,\delta,\uparrow> 
\left(a\,\cos(\frac{\delta}{2})|2,\downarrow>
+b\,\sin(\frac{\delta}{2})|2,\uparrow>\right)\mbox{\hspace{0.5cm}}\nonumber\\
+|m_-> |1,\delta,\downarrow> 
\left(a\,\sin(\frac{\delta}{2})|2,\downarrow>
-b\,\cos(\frac{\delta}{2})|2,\uparrow>\right)\quad.
\label{u9}
\end{eqnarray}
According to the Copenhagen interpretation one ought to 
apply the concept of the reduction of the wave function,
which yields that the state of the second, {\em undisturbed}
particle has the state (after the measurement)
\begin{eqnarray}
|\varphi_+^{(2)}>=\mbox{\hspace{11cm}}\nonumber\\
\left(|a|^2\cos^2(\frac{\delta}{2})+|b|^2\sin^2(\frac{\delta}{2})\right)^
{-\frac{1}{2}}
\left(a\,\cos(\frac{\delta}{2})|2,\downarrow>
+b\,\sin(\frac{\delta}{2})|2,\uparrow>\right)\quad,
\label{u10}
\end{eqnarray}
if we have measured $\frac{\hbar}{2}$ and
\begin{eqnarray}
|\varphi_-^{(2)}>=\mbox{\hspace{11cm}}\nonumber\\
\left(|a|^2\sin^2(\frac{\delta}{2})+|b|^2\cos^2(\frac{\delta}{2})\right)^
{-\frac{1}{2}}
\left(a\,\sin(\frac{\delta}{2})|2,\downarrow>
-b\,\cos(\frac{\delta}{2})|2,\uparrow>\right)\quad,
\label{u11}
\end{eqnarray}
if we have measured $-\frac{\hbar}{2}$. These states depend on
$\delta$, i.e., on the quantity $\hat S_{1z'}$ 
which has been measured on the
first particle.

According to the present theory there is no collapse of the
wave function, however, the states (\ref{u10}) and (\ref{u11}) may be
easily identified by $\hat \rho_{P_2}(P_1+P_2)$. 

Indeed, 
\begin{eqnarray}
\hat \rho_{P_1+P_2}(P_1+P_2+M)\mbox{\hspace{8cm}}\nonumber\\
=|1,\delta,\uparrow>|\varphi_+^{(2)}> 
\left(|a|^2\cos^2(\frac{\delta}{2})+|b|^2\sin^2(\frac{\delta}{2})\right)
<\varphi_+^{(2)}|<1,\delta,\uparrow|\mbox{\hspace{0.5cm}}\nonumber\\
+|1,\delta,\downarrow>|\varphi_-^{(2)}> 
\left(|a|^2\sin^2(\frac{\delta}{2})+|b|^2\cos^2(\frac{\delta}{2})\right)
<\varphi_-^{(2)}|<1,\delta,\downarrow|\;.\nonumber
\end{eqnarray}
Its eigenstates are
\begin{eqnarray}
|1,\delta,\uparrow>|\varphi_+^{(2)}> \label{u12}
\end{eqnarray}
and
\begin{eqnarray}
|1,\delta,\downarrow>|\varphi_-^{(2)}>\;.\label{u13}
\end{eqnarray}
Thus $\hat
\rho_{P_1+P_2}(P_1+P_2)=|\psi_{P_1+P_2}><\psi_{P_1+P_2}|$ 
where $|\psi_{P_1+P_2}>$ coincides with either (\ref{u12}) or
 (\ref{u13}). Calculating $\hat
\rho_{P_2}(P_1+P_2)=Tr_{P_1}\left(\hat
\rho_{P_1+P_2}(P_1+P_2)\right)$ we arrive at the expressions $|\varphi_+^{(2)}><\varphi_+^{(2)}$ or
$|\varphi_-^{(2)}><\varphi_-^{(2)}|$ (cf. Eqs.(\ref{u10}),(\ref{u11}), respectively). 
The dependence on $\delta$ is now
easily understood:
the measurement done on $P_1$ influences the quantum reference
system $P_1+P_2$, hence $\hat \rho_{P_2}(P_1+P_2)$ 
depends on the measurement, although $P_2$ has not been
influenced. Certainly, Einstein separability is not violated.

Einstein separability requires now
that $\hat \rho_{P_2}(P_2)$ must be independent of the
measurement done on $P_1$. Direct calculation shows
that $\hat \rho_{P_2}(P_1+P_2+M)$ is the same both before
and after the measurement, thus its eigenstates are unchanged,
too. Therefore, $\hat \rho_{P_2}(P_2)$ is not influenced 
by the measurement. One may also prove\cite{Bene} that quite
generally, if the system $A$ is  separated from the systems 
$B$ and $C$, $\hat \rho_A(A)$ is independent of the
interaction between $B$ and $C$. 

\section{Explanation of the violation of Bell's inequality}

Let us consider again the previous two-particle system. 
In order to exhibit
 the mathematical structure we write the state (\ref{u6}) as
\begin{eqnarray}
\sum_j c_j
|\phi_{P_1,j}>|\phi_{P_2,j}>\label{u14}
\end{eqnarray}
where $c_1=a$, $c_2=-b$, $
|\phi_{P_1,1}>=|1,\uparrow>$, $
|\phi_{P_1,2}>=|1,\downarrow>$, $
|\phi_{P_2,1}>=|2,\downarrow>$, $
|\phi_{P_2,2}>=|2,\uparrow>$.
When the two particle system is in the state (\ref{u14}),
there are strong correlations
between the states $\hat \rho_{P_1}(P_1)=|\psi_{P_1}><\psi_{P_1}|$ 
and $\hat \rho_{P_2}(P_2)=|\psi_{P_2}><\psi_{P_2}|$. 
Provided that the system $P_1+P_2$
is initially isolated, applying {\bf Postulate 9} we obtain that the
probability that $|\psi_{P_1}>=|\phi_{P_1,j}>$ and 
$|\psi_{P_2}>=|\phi_{P_2,k}>$ is
$
P(P_1,j,P_2,k)=|c_j|^2\delta_{j,k}
$.

Let us consider now a typical experimental situation, 
when measurements on both
particles are performed. We shall show
that according to the present theory the observed
correlations are exclusively due to the previous interaction
between the particles. Before the measurements the internal
state of the isolated system $P_1+M_1+P_2+M_2$ ($P_1,P_2$
stands for the particles and $M_1,M_2$ for the measuring
devices, respectively) is given by\hfill\break
$
\left(\sum_j c_j
|\phi_{P_1,j}>|\phi_{P_2,j}>\right)|m^{(1)}_0> |m^{(2)}_0>$,
while it is
\begin{eqnarray}
\sum_j c_j
\hat U_t(P_1+M_1)\left(|\phi_{P_1,j}>|m^{(1)}_0>\right)
\hat U_t(P_2+M_2)\left(|\phi_{P_2,j}>|m^{(2)}_0>\right) \quad,
\label{u15}
\end{eqnarray}
a time $t$ later, i.e. during and after the measurements. Here 
$\hat U_t(P_i+M_i)$ ($i=1,2$) stands for the unitary time evolution operator
of the closed system $P_i+M_i$.

Eq.(\ref{u15}) implies that the internal states of 
the closed systems $P_1+M_1$ and $P_2+M_2$
evolve unitarily and do not influence each
other. This time evolution can be given explicitly through the 
relations
\begin{eqnarray}
|\xi(P_i,j)>|m^{(i)}_0>\;\rightarrow \;|\xi(P_i,j)>|m^{(i)}_j>\quad,
\label{u16}
\end{eqnarray}
where $i,j=1,2$ and $|\xi(P_i,j)>$ is the $j$-th eigenstate of the 
spin measured 
on the $i$-th particle along an axis $z^{(i)}$ which closes an angle
$\vartheta_i$ with the original $z$ direction. The time evolution of 
the internal state of the closed systems $P_i+M_i$ is given explicitly by
$
|\psi_{P_i}>|m_0^{(i)}>\;\rightarrow \;
\sum_j <\phi_{P_i,j}|\psi_{P_i}>|\phi_{P_i,j}>|m_j^{(i)}>
$.
As we see, the $i$-th measurement process 
is completely determined by the initial internal states of the
particle $P_i$. Therefore, any correlation between
the measurements may only stem from the initial correlation
of the internal states of the particles.

For the calculation of the state $\hat \rho_{M_1}(M_1)$ (which corresponds to the
measured value) one needs to know the state of the whole isolated system
$P_1+P_2+M_1+M_2$.
Using Eq.(\ref{u16}) the final state (\ref{u15}) may be written as
\begin{eqnarray}
\sum_{j,k}\left(
\sum_l c_l<\xi(P_1,j)|\phi_{P_1,l}><\xi(P_2,k)|\phi_{P_2,l}>
\right)\mbox{\hspace{4cm}}\nonumber\\
\times |m^{(1)}_j>|m^{(2)}_k>|\xi(P_1,j)>|\xi(P_2,k)>.\nonumber
\end{eqnarray}

Direct calculation shows that
\begin{eqnarray}
\hat \rho_{M_1}(P_1+P_2+M_1+M_2)\mbox{\hspace{5cm}}\nonumber\\
=\sum_j\left(\sum_l
|c_l|^2 |<\xi(P_1,j)|\phi_{P_1,l}>|^2\right)|m^{(1)}_j><m^{(1)}_j|
.\nonumber
\end{eqnarray}
Note that it is independent of the second measurement.

According to {\bf Postulate 6}  $|\psi_{M_1}>$ is one of the
$|m^{(1)}_j>$-s. (Similarly one may derive that
$|\psi_{M_2}>$ is one of the
$|m^{(2)}_k>$-s.)
The probability of the observation of the $j$-th result (up or
down spin in a chosen direction) is
\begin{eqnarray}
P(M_1,j)=\sum_l |c_l|^2 |<\xi(P_1,j)|\phi_{P_1,l}>|^2\;.\label{u17}
\end{eqnarray}
This may be interpreted in conventional terms: $|c_l|^2$ is
the probability that $|\psi_{P_1}>=|\phi_{P_1,l}>$,
and $|<\xi(P_1,j)|\phi_{P_1,l}>|^2$ is the conditional
probability that one gets the $j$-th result if $|\psi_{P_1}>=|\phi_{P_1,l}>$.
Thus we see that the initial internal state of $P_1$ determines
the outcome of the first measurement in the usual 
probabilistic sense.  One may show quite similarly that
the initial internal state of $P_2$ determines
the outcome of the second measurement in the same way.

But doesn't it mean that the internal states of $P_1$ and $P_2$
play the role of local hidden variables? Not at all, because hidden variables
are thought to be comparable with the results of the measurements
so that their joint probability may be defined, while in our theory
there is no way to define the joint probability
$P(P_1,l_1,P_2,l_2,(0);M_1,j,M_2,k,(t))$, i.e., the probability that initially
$|\psi_{P_1}>=|\phi_{P_1,l_1}>$ and $|\psi_{P_2}>=|\phi_{P_2,l_2}>$
{\em and} finally $|\psi_{M_1}>=|m^{(1)}_j>$ 
and $|\psi_{M_2}>=|m^{(2)}_k>$. Intuitively we would write
\begin{eqnarray}
P(P_1,l_1,P_2,l_2,(0);M_1,j,M_2,k,(t))\mbox{\hspace{4cm}}\nonumber\\
=|c_{l_1}|^2\delta_{l_1,l_2}|<\xi(P_1,j)|\phi_{P_1,l_1}>|^2
|<\xi(P_2,k)|\phi_{P_2,l_2}>|^2\;,\label{u18}
\end{eqnarray}
as $|c_{l_1}|^2\delta_{l_1,l_2}$ is
the joint probability that $|\psi_{P_1}>=|\phi_{P_1,l}>$ 
and $|\psi_{P_2}>=|\phi_{P_2,l}>$, and $|<\xi(P_i,j)|\phi_{P_i,l_i}>|^2$ is the conditional
probability that one gets the $j$-th result in the $i-th$
measurement if initially $|\psi_{P_i}>=|\phi_{P_i,l_i}>$ ($i=1,2$).
Certainly the existence of such a joint probability would immediately imply the
validity of Bell's inequality, thus it is absolutely important 
to understand why this probability does not exist.

Let us mention, first of all, that using  {\bf  Postulate 9} for $n=2$, we may calculate the correlation
between the measurements, i.e., the joint probability 
that $|\psi_{M_1}>=|m^{(1)}_j>$ {\em and} $|\psi_{M_2}>=|m^{(2)}_k>$.
We obtain
\begin{eqnarray}
P(M_1,j,M_2,k)
=\left|\sum_l c_l<\xi(P_1,j)|\phi_{P_1,l}><\xi(P_2,k)|\phi_{P_2,l}>\right|^2
\quad.\label{u19}
\end{eqnarray}
This is the usual quantum mechanical expression 
which violates Bell's inequality and whose correctness is experimentally
proven. Thus our theory gives the correct expression for the correlation.
Nevertheless, if the joint probability 
(\ref{u18}) exists, it leads to 
\begin{eqnarray}
P(M_1,j,M_2,k)=
\sum_l |c_l|^2 |<\xi(P_1,j)|\phi_{P_1,l}>|^2
|<\xi(P_2,k)|\phi_{P_2,l}>|^2\label{u20}
\end{eqnarray}
 which satisfies Bell's inequality and contradicts
Eq.(\ref{u19}). Let us demonstrate that no such contradiction appears.

Evidently, the joint probability 
$P(P_1,l_1,P_2,l_2,(0);M_1,j,M_2,k,(t))$ can be physically meaningful 
only if one can compare the initial internal states of $P_1$ and $P_2$
with the final internal states of $M_1$ and $M_2$ by suitable
nondisturbing measurements. It turns out, however, that any attempt
for such a comparison influences the system so strongly
that the correlations $P(M_1,j,M_2,k)$ change.

If we try to compare the initial internal 
states of $P_1$ and of $P_2$ with the final
internal states of $M_1$ and $M_2$, 
the first difficulty appears because we want to compare states
given at different times. Nevertheless, as the initial internal state
of $P_i$ is uniquely related to the final internal state of the
system $P_i+M_i$, the joint probability $P(P_1,l_1,P_2,l_2,(0);M_1,j,M_2,k,(t))$ 
(if exist) coincides with $P(P_1+M_1,l_1,P_2+M_2,l_2,M_1,j,M_2,k)$,
where all the occuring states are given after the measurements.
As the systems $P_1+M_1,\;P_2+M_2,\;M_1,\;M_2$ are not disjointed, 
our {\bf Postulates} do not provide us with an expression for
the joint probability we are seeking for.
 If we check $|\psi_{M_1}>$ and $|\psi_{M_2}>$ by
 suitable nondisturbing measurements, we destroy
 $|\psi_{P_1+M_1}>$ and $|\psi_{P_2+M_2}>$ (cf. the discussion
 at the end of Section 2.), inhibiting any comparison.
  On the other hand, if we check
 first $|\psi_{P_1+M_1}>$ and $|\psi_{P_2+M_2}>$, then 
$P(M_1,j,M_2,k)$ changes.\footnote{
Therefore, in the original situation the
joint probability $P(P_1+M_1,l_1,P_2+M_2,l_2,M_1,j,M_2,k)$ cannot be defined at all.
Here we are faced with the feature of our theory discussed at the 
end of Section 2: although states with respect to different quantum reference
systems do exist individually, their simultaneous occurence or comparatibility
may not be defined.}  Indeed, after suitable measurements 
performed on
$P_i+M_i$ (which do not change the internal states of $P_i+M_i$)\footnote{
This is equivalent by recording the initial internal state
of $P_i$.} by further measuring devices $\tilde M_i$ we get for the
internal state  of the whole system
 \begin{eqnarray}
\sum_l c_l\left( \sum_j <\xi(P_1,j)|\phi_{P_1,l}>|\xi(P_1,j)>|m^{(1)}_j>\right)
\mbox{\hspace{2.5cm}}\nonumber\\
\times\left( \sum_k <\xi(P_2,k)|\phi_{P_2,l}>|\xi(P_2,k)>|m^{(2)}_k>
\right)|\tilde m^{(1)}_l>|\tilde m^{(2)}_l>\;.\label{u21}
\end{eqnarray} 
As the systems $M_1,\;M_2,\;\tilde M_1,\;\tilde M_2$ are disjointed,
we may apply {\bf Postulate 9} for $n=4$ and we indeed get for 
$P(\tilde M_1,l_1,\tilde M_2,l_2,M_1,j,M_2,k)$ the expression
(\ref{u18}). Do we get then a contradiction with Eq.(\ref{u19})?
No, because applying {\bf Postulate 9} for $n=2$ directly, we get in this case
Eq.(\ref{u20}) instead of Eq.(\ref{u19}). Thus we see that the
extra measurements have changed the correlations and our theory 
gives account of this effect consistently.

Summarizing, we have seen that the initial internal state
of $P_1$ ($P_2$) determines the first (second) measurement
process, therefore, these states 'carry' the initial correlations
and 'transfer' them to the measuring devices. 
As the measurement processes do not influence each other, 
the observed correlations may stem only from the 'common past'
of the particles.
On the other hand, any attempt to
compare the initial internal states of $P_1$ and $P_2$ with
the results of both measurements changes the correlations,
thus a joint probability for the simultaneous existence of these states
cannot be defined. This means that the reason for the
 violation of Bell's inequality is that the usual derivations
 always assume that the states (or 'stable properties')
 which carry the initial correlations can be freely compared with the results
 of the measurements. This comparability is usually 
 thought to be a consequence of realism.
 According to the present theory, the above assumption
 goes beyond the requirements of realism and proves to be wrong,
 because each of the states $|\psi_{P_1+M_1}>$, 
$|\psi_{P_2+M_2}>$, $|\psi_{M_1}>$ and $|\psi_{M_2}>$ exists individually,
but they cannot be compared without changing the correlations.
\vskip0.5cm
In conclusion, it has been shown that by suitable redefinition of the
physical meaning of the quantum states and their interrelations 
one can solve the EPR paradox
and can explain the violation of Bell's inequality without giving up
realism, inductive inference or Einstein separability. 

\section{Acknowledgements}

The author is indebted to A.Bringer, G.Eilenberger, M.Eisele, 
R.Graham, G.Gy\"orgyi, F.Haake,Z.Kaufmann, H.Lustfeld, P.Rosenqvist,
 P.Sz\'epfalusy, G.Tichy and G.Vattay for useful discussions 
 and remarks,
 to P.Sz\'epfalusy also for his continued interest in 
 the work and for encouragement, and to G.Tichy for correcting several
 grammatical errors in the manuscript.
 The author wants to thank for the hospitality of the {\em Institut f\"ur 
 Festk\"orperphysik, Forschungszentrum J\"ulich GmbH} where a 
 substantial 
 part of the work has been done. 
 This work has been partially supported by the Hungarian Aca\-demy of 
 Sciences
 under Grant Nos. OTKA T 017493, OTKA F 17166 and OTKA F 019266.
The present paper came into being 
 within the framework of a scientific and technological cooperation agreement
 between the Hungarian and the German government, 
 as a result of a research cooperation supported
 by the OMFB (Hungary) and the BMFT (Germany).

\end{document}